\documentclass[useAMS,usenatbib,usegraphicx]{mn2e}

\usepackage{aas_macros}
\usepackage{multirow}
\title[Satellite Number Density Profiles]{
Satellite Galaxy Number Density Profiles in the Sloan Digital Sky Survey}
\author[Guo et al]{Quan Guo, Shaun Cole, Vincent Eke, Carlos Frenk\\
Institute for Computational Cosmology, Department of Physics, 
Durham University, Science
Laboratories, South Rd, Durham DH1 3LE.} 

\newcommand{\satmc}{M_{\rm{c}}}
\newcommand{\satdmbin}{\Delta M_{\rm{bin}}}
\newcommand{\satdmf}{\Delta M_{\rm{faint}}}
\newcommand{\satdzs}{\Delta z_{\rm{s}}}
\newcommand{\satap}{\alpha_{\rm{p}}}

\newcommand{\ri}{R_{\rm{inner}}}
\newcommand{\ro}{R_{\rm{outer}}}
\newcommand{\aii}{A^{\rm{ann}}_{ij}}
\newcommand{\ai}{A^{\rm{inner}}_{i}}
\newcommand{\aio}{A^{\rm{outer}}_i}
\newcommand{\msatt}{M_{\rm sat}^{\rm trun}}
\newcommand{\lii}{N^{\rm{inner}}_i(r^{\rm ann}_j)}
\newcommand{\lio}{N^{\rm{outer}}_i}
\newcommand{\lsat}{\Sigma^{\rm{sat}}_i(r^{\rm ann}_j)}

\newcommand{\emlsat}{\widetilde{\Sigma}^{\rm{sat}}(r_j^{\rm ann},\msatt)}

\newcommand{\lcdm}{$\rm{\Lambda CDM}$}

\newcommand\lsim{\mathrel{\hbox{\rlap{\hbox{\lower4pt\hbox{$\sim$}}}\hbox{$<$}}}}
\begin{document}


\maketitle

\label{firstpage}

\begin{abstract}
We study the spatial distribution of satellite galaxies around
isolated primaries using the Sloan Digital Sky Survey (SDSS)
spectroscopic and photometric galaxy catalogues. We select isolated
primaries from the spectroscopic sample and search for potential
satellites in the much deeper photometric sample.  For specific
luminosity primaries we obtain robust statistical results by stacking
as many as $\sim\negthinspace 50,000$ galaxy systems.  We find no
evidence for any anistropy in the satellite galaxy distribution
relative to the major axes of the primaries.  We derive accurate
projected number density profiles of satellites down to 4~magnitudes
fainter than their primaries.  We find the normalized satellite
profiles generally have a universal form and can be well fitted by
projected NFW profiles.  The NFW concentration parameter increases
with decreasing satellite luminosity while being independent of the
luminosity of the primary except for very bright primaries.  The
profiles of the faintest satellites show deviations from the NFW form
with an excess at small galactocentric projected distances.  In
addition, we quantify how the radial distribution of satellites
depends on the colour of the satellites and on the colour and
concentration of their primaries.
\end{abstract}

\begin{keywords}
Galaxies: dwarf, Galaxies: structure, Galaxies: Local Group, Galaxies: fundamental parameters

\end{keywords}

\section{Introduction}
In the \lcdm~model, smaller structures falling into larger haloes can
survive there as substructures and host observed satellite
galaxies. Therefore, the distribution of satellites around primaries
holds important information about galaxy formation, the population of
substructures and even the nature of dark matter. In the past decade
or so, fainter satellites around the Milky Way (MW) and M31 have been
discovered in the Sloan Digital Sky Survey (SDSS) \citep[e.g.][]{
gre00,van00,zuc04,willman05,zuc06,zuc07,mar06,mar08,irw07,sim07,bel08,liu08,wat09,sla11},
which have proven to be important observational data widely
used to constrain the cosmological model
\citep[e.g.][]{kly02,lovell11}.  In addition, these data can
constrain attempts
to understand the formation of galaxies in subhalos using 
semi-analytic modelling techniques
\citep[e.g.][]{ben02,koposov09,munoz09,busha10,cooper10,mac10,li10,font11,wang12}
and full N-body/gasdynamic simulations to investigate the physics of
satellite galaxies \citep[e.g.][]{libeskind07,okamotof09,oka10,wad10,par11}.
%can be constrained by these data.

The large body of work on satellite galaxies reflects the fact that
they are not only a critical small scale test of the \lcdm\ model, but
also a probe of the nature of dark matter; yet the satellite data
with which all theories and models are compared with are merely those
of two primaries, the MW and M31.  Although the satellite populations
of MW and M31 are known better than other satellite systems,
there is no guarantee that they are typical. Clearly, robust and
reliable tests of cosmological and galaxy formation models require
comparison with a statistically representative sample of galaxies and
their satellites.

Early studies were limited by the relatively small satellite samples available
at the time \citep{hol69,lor94,zar93,zar97b}. 
With the advent of large galaxy redshift surveys such as
the 2dF Galaxy Redshift Survey \citep[2dFGRS;][]{col01} and the SDSS 
\citep{yok00}, it is now possible to
construct external galaxy samples spanning a much larger volume.
Studies with significantly improved statistics have been carried out
using these new surveys. For example, \cite{sal04} studied 
the spatial distribution of satellites around primaries using the
2dFGRS. More recently \cite{yan06} studied how 
spectroscopically identified satellite galaxies were distributed in SDSS groups
relative to the orientation of the central galaxy.
However, due to the flux limit of
redshift surveys, analyzing the satellite systems of external isolated
galaxies is still challenging because typically 
only one or two satellites are detected
per primary galaxy. In addition, the
real space position of satellites with respect to their primaries is
uncertain due to redshift space distortions and projection effects.
To circumvent the aforementioned problems, we \citep[][hereafter Paper
I]{gu11} have developed a method of stacking the primaries and their
satellites in order to obtain a fair and complete sample that can
yield statistically robust results for certain classes of primary
galaxies.  This method has been successfully applied to the estimation
of the satellite luminosity functions of isolated primary galaxies in the
SDSS.

In this work, for the same primary and satellite samples we explored
in Paper I, we are now interested in the average spatial profile of
the distribution of these satellites around their primaries. These
density profiles are an important tracer of the distribution of
substructures in the primary halo and can provide us with useful
information to test current models of the formation and evolution of
dark matter haloes.

In the cold dark matter (CDM) cosmological model, the density profiles
of dark matter haloes follow a universal form \cite[hereafter NFW
profiles]{nfw96,nfw97} with an inner cusp, $\rho (r)\propto r^{-1}$,
and an outer slope of $\rho (r)\propto r^{-3}$. The transition scale,
$r_{\rm s}$, is normally specified through the concentration,
$c=r_{200}/r_{\rm s}$, where $r_{200}$ is defined as the radius
enclosing a mean interior density 200 times the critical density.
Besides the overall mass profile, it is remarkable that the spatial
distribution of dark matter substructures, which could host satellites
galaxies, also follows this universal form independent of the mass of
the substructures \citep{die04,spr08,lud09}.  However, the number of
observed satellite galaxies around the MW and M31 is much smaller than
the number of substructures predicted by \lcdm~\citep{kly99,moo99},
giving rise to the so-called ``missing satellites problem''.
Statistically robust number density profiles of observed satellites
will certainly help us understand how satellite galaxies populate the
substructures.  In addition, a reliable density profile is required to
extrapolate the incomplete observational data of satellites around the
MW and M31 to compare with models \citep[e.g. ][]{kop08,tol08}.

The recognition of the importance of the spatial profiles of systems
of satellite galaxies has resulted in many studies. Early work with
samples from a limited volume have mainly focused on fitting the slope
of the density profile of satellite galaxies around isolated primaries
\citep{lak80,vad91,lor94}.  With large galaxy redshift surveys, the
dependence of the profiles on the colour and morphology of primaries
has begun to be explored \citep[e.g.][]{sal05,chen06}.  \cite{kly09}
studied the projected number density profiles and velocity dispersion
around isolated red primaries using the SDSS redshift sample.
\cite{mor09} used an iterative method, tested on mock galaxy
catalogues, to find satellite systems around central galaxies with a
range of luminosities in the SDSS. The distribution of velocities of
the satellites was used to infer mass-to-light ratios as a
function of central galaxy luminosity.  Closely related to this are
studies of the radial distributions of satellite galaxies in clusters,
groups \citep{li07,wang11} and on smaller scales \citep{wat12}.
Further work has focused on the distribution of satellites around
intermediate redshift galaxies \citep{nie11}, elliptical primaries
\citep{smi04} and isolated galaxies in the SDSS \citep{lar11}.
Besides the studies that statistically estimate mean number density
profiles, \citet{kim11} have directly measured the number density
profile of the nearby field galaxy M106.

In this paper, we are not only interested in the projected number
density profile of satellites around isolated primary galaxies binned by 
luminosity, colour and morphology, but also on the dependence of the
profiles on the properties of the satellites themselves. To this end, we
select our primary samples from the SDSS DR8 spectroscopic sample
($\sim 660~000$ galaxies) and satellite candidates from the
photometric samples ($\sim 96~000~000$ galaxies) with the same criteria
as in Paper I to build significantly large samples.  We restrict
the photometric sample to galaxies brighter than $m_r=20.5$ as in
Paper I (See Paper I, Section 4) to ensure completeness.  Based on
these large samples, we explore the dependence of the density profiles
on the properties of primaries and satellites.

The remainder of this paper is organised as follows. In Section 2,
we briefly describe the selection of primary galaxies and their
satellites; in Section 3, we develop a method for estimating 
the projected satellite number density profile; in Section 4,
we present our estimate of the projected satellite number density
for different primary samples. We conclude, in Section 5, 
with a summary and discussion of our results.
Throughout the paper we assume a fiducial ${\rm\Lambda CDM}$ cosmological model
with $\Omega_M=0.3$, $\Omega_{\Lambda}=0.7$ and $H_0 = 70$km~s$^{-1}$~Mpc$^{-1}$.

%1). motivation
%faint galaxies are tell us about galaxy formation and theory of DM
%the satellite of Milkway is not not enough, roughly servy is only one or tw atellites
%Sloan survey advagntage
%density plot is imporat 
  %tell use how population of dark matter
  %to extroplate of odat. which can't doign in ML.
%subhalos in Lbamd, only smalle of substrucat will hold visual galaxies, help us to under stand.
%2) discribe previous work
%3) why you are doing, what you get so far

\section{Sample and Method}

\subsection{Data and sample selection}
  We have built two separate catalogues similar to those in Paper~I.
  The smaller one is of galaxies from the main SDSS spectroscopic
  catalogue from which we select our primary galaxies (hereafter the
  spectroscopic catalogue). The larger one is of galaxies with
  photometric redshifts and magnitudes from which we select the
  neighbouring galaxies (hereafter the photometric catalogue).  The
  spectroscopic catalogue is constructed from the SDSS DR8
  spectroscopic subsample (north galactic cap) including all objects
  with high quality redshifts (zconf $> 0.7$ and specClass $=2$) and
  a Petrosian magnitude $r \le 17.77$.  The photometric galaxy
  catalogue is from the SDSS DR8 photometric subsample (north
  galactic cap) and includes only objects that have photometric
  redshifts, none of the flags BRIGHT, SATURATED, or SATUR\_CENTER\footnote{When applying our isolation criteria to reject primaries with bright 
neighbours we use a source catalogue that also includes objects for which
SATURATED and/or SATUR\_CENTER flags are set. These objects are mainly
stars and we prefer to reject systems contaminated by bright stars
as the presence of such unmasked stars could effect the efficiency 
with which background galaxies are detected.}
  set and model magnitudes $r \le 22.0$. We select only objects with
  corresponding entries in the SDSS database PhotoZ table, which
  naturally selects galaxies and excludes stars.  As galaxies with
  $r \le 17.77$ are included in both SDSS catalogues, a small
  fraction of the photometric catalogue galaxies also have
  spectroscopic redshifts.  We use de-reddened model $ugriz$
  magnitudes and $k$-correct all galaxies to $z=0$ with the IDL code
  of \citet{bla07}. We estimate $V$-band magnitudes
  from $g$ and $r$-band magnitudes assuming
  $V=g-0.55(g-r)-0.03$ \citep{smi02} and all our sample selection
  and magnitude cuts
  are performed using this $V$-band magnitude.

  \begin{figure}
     \includegraphics[width=84mm]{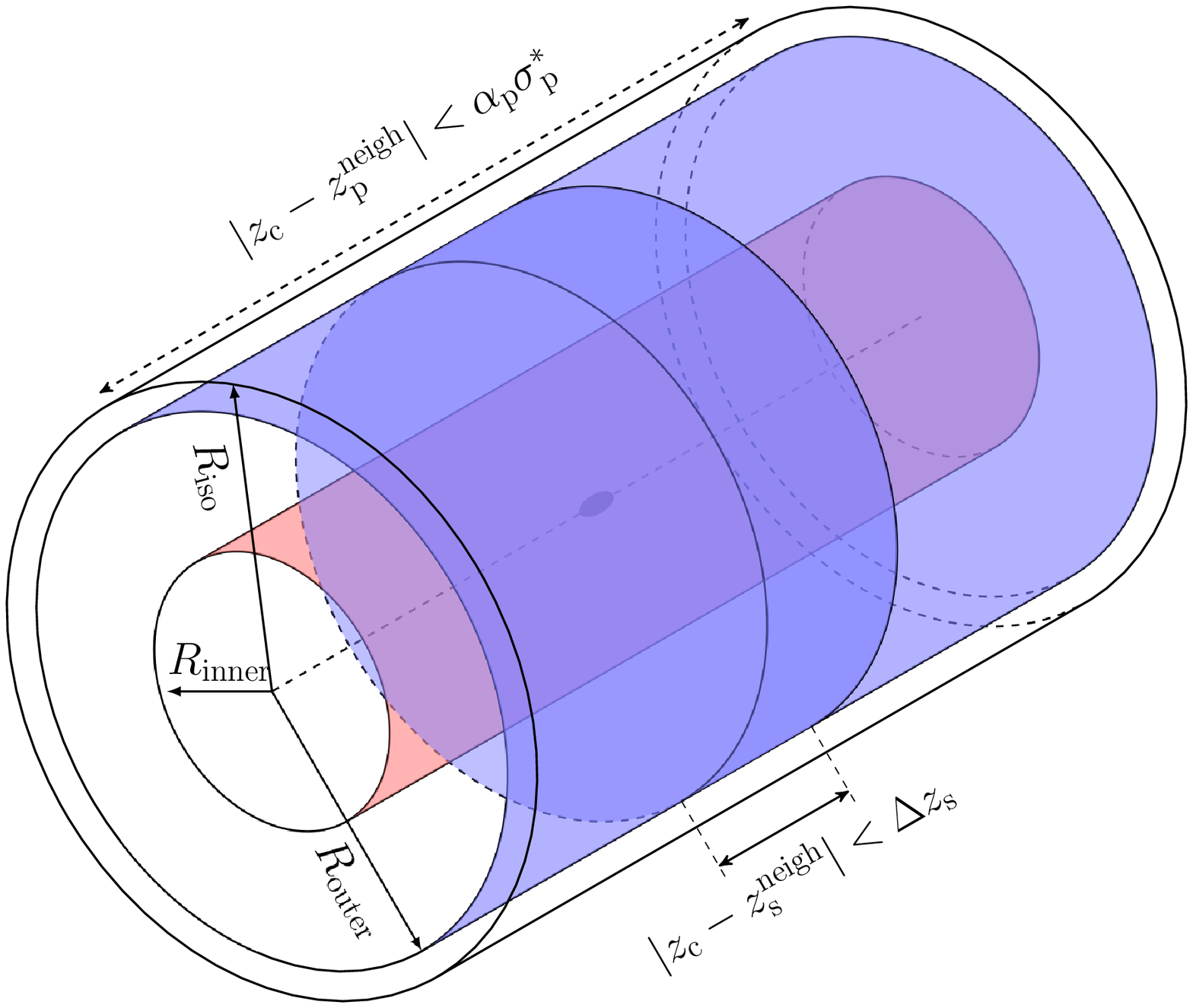}
     \caption{Schematic diagram of the sample selection procedure.
        For each acceptable primary, there should be no neighbouring,
        sufficiently bright, galaxies within a cylinder of
        radius $R_{\rm iso}$, centred on the primary, and nearby in
        redshift space. As defined 
        in Fig.~1 in Paper I, nearby means either
	$|z_{\rm c}-z_{\rm s}^{\rm neigh}| < \Delta z_{\rm s}$,
        where $\Delta z_{\rm s}$
        is the maximum allowed spectroscopic redshift difference between a
        primary ($\rm c$) and another galaxy ($\rm s$), or
	$|z_{\rm c}-z_{\rm p}^{\rm neigh}|<\alpha_{\rm p}\sigma_{\rm p}^*$,
	where $\sigma_{\rm p}$ is the measurement error of the
        photometric redshift and $\alpha_{\rm p}$ is the tolerance of
        the error, for galaxies that have no spectroscopic redshift.
	Satellites will lie nearby in redshift space and within the
        cylinder of radius $\ri$ (red), whereas the local background
        to be subtracted is determined from the volume between this
        inner cylinder and the outer one with radius $\ro$.
        }
        \label{fig:method}
 \end{figure}

 Our sample of isolated primary galaxies is chosen using the same
 criteria as in Paper~I, illustrated in Fig.~\ref{fig:method}. We select primary galaxy candidates of
 absolute magnitude, $M_{\rm p}$, in the range $M_{\rm{C}} -\Delta
 M_{\rm{bin}} < M_{\rm p} \le M_{\rm{C}} +\Delta M_{\rm{bin}}$. We
 then filter these primary candidates, using a series of criteria
 summarised in figure~1 of Paper~I, to guarantee that a) there are no
 luminous neighbouring galaxies projected within $2R_{\rm inner}$ of
 the primary, unless these luminous neighbours are sufficiently
 separated in redshift from the primary and appear here due to a
 chance projection; b) the satellite search areas (projected distance
 $R_{\rm inner}$ from the primary) around each primary do not overlap
 with each other.  Further details of the generation of the two
 samples are may be found in Paper~I.  The values of the selection
 parameters, $\{\satdmbin,\satdmf, \satdzs,\satap, m_{\rm v}^{\rm
   lim}\}= \{0.5, 0.5, 0.002, 2.5,20.5\}$, are the same as the default
 values in Paper~I. Here $\satdmbin$ is the half-width of the primary
 magnitude $\satmc$ bin, $\satdmf$ is the magnitude difference between
 the primaries and satellites used to isolate primaries, $\satdzs$ and
 $\satap$ are the parameters used to exclude galaxies that are at a
 significantly different spectroscopic redshift and photometric
 redshift respectively.  The meaning of these parameters is explained
 in Fig.~\ref{fig:method} (see section~2 of Paper~I for more
 details). One small change relative to Paper~1 is that $\ri$ is
 chosen to increase with increasing primary luminosity, in order to
 ensure that no satellites are missed for the most luminous primaries.

  \subsection{Estimating the projected satellite number density profile}
  \begin{figure*}
     \includegraphics[width=126mm]{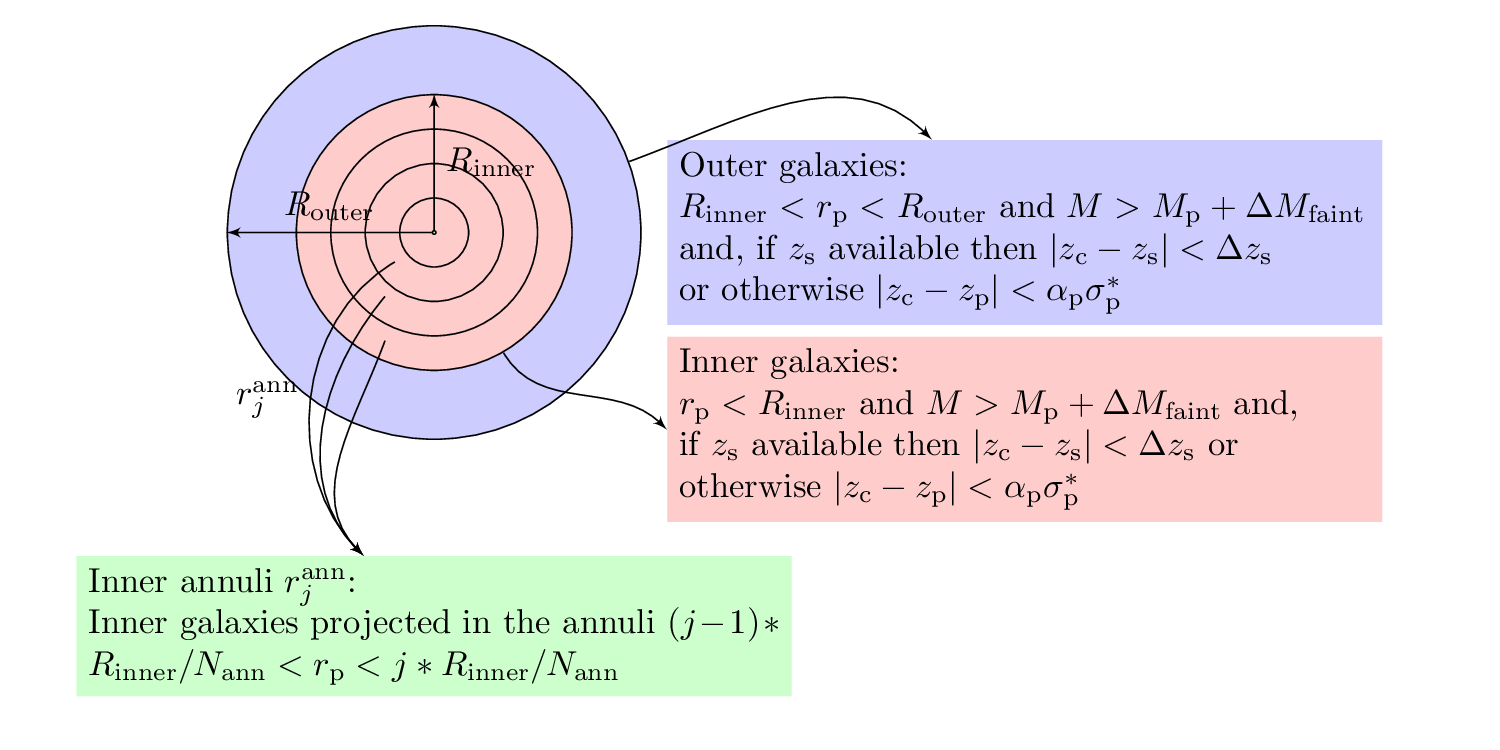}
     \caption{Schematic showing the selection of potential satellite galaxies
       in annuli of radii $r^{\rm ann}_j$ within $\ri$ and 
       a reference sample bounded by the radii
       $\ri<r<R_{\rm outer}$, used to subtract the residual
       contaminating background. For both samples we apply the stated
	 redshift cuts to reduce background contamination. We also apply
	 the stated absolute magnitude cut to both samples (assuming the
	 neighbouring galaxies are at the same redshift as the primary)
	 though this cut is redundant unless
	 $R_{\rm{outer}}>2\ri$ as otherwise the existence of
	 such bright neighbouring galaxies would automatically lead to the
	 exclusion of the primary galaxy. }
     \label{fig:sat_pro}
 \end{figure*}

Once our primary galaxies are defined, their potential satellites are
found from the photometric galaxy catalogue as depicted in
Fig.~\ref{fig:sat_pro}. The method we use is similar to that in
Paper~I, except that for each primary, the number of galaxies is counted
and binned by their projected radial distances from the primary,
$N(r^{\rm ann}_j)$, as well as by their luminosity.  That is, for the
$i$th primary galaxy, the number of inner galaxies in each annulus,
$\lii$, is found by counting all the neighbouring galaxies within the
annulus of radius $r_j^{\rm ann}$ that satisfy the following
conditions: at least $\satdmf$ fainter than the primary; if they have
a spectroscopic redshift, $z_{\rm{s}}$, then they should satisfy
$|z_{\rm{c}}-z_{\rm{s}}|<\satdzs$; or if they only have a photometric
redshift $z_{\rm{p}}$, then they  should satisfy
$|z_{\rm{c}}-z_{\rm{p}}|<\alpha_{\rm{p}}\sigma_{\rm{p}}^*$, where
$\sigma_{\rm{p}}^*$ is the error in the photometric redshift as
defined in Paper~1. The number of outer galaxies, $\lio$, is
determined by applying the same conditions to galaxies in the outer
area between $\ri$ and $\ro$.  Assuming, for now, that few genuine
satellites will be projected beyond $\ri$ we can estimate the surface
density of genuine satellites in each annulus as 
  \begin{equation}
	\lsat=\frac{\lii}{\aii}-\frac{\lio}{\aio},
  \end{equation}
  where $\aii$ and $\aio$ are the areas of the inner annulus and outer
  region respectively. If necessary, we take account of the sky
  coverage of SDSS DR8 by reducing the areas $\aii$ and $\aio$ by the
  amounts defined by the DR7 mask described in \citet{norb11}.

Because of the apparent magnitude limit of SDSS, which we take to be
$m^{\rm lim}= 20.5$, we count the faintest satellites only at low
redshift and are progressively limited to more and more luminous
satellites with increasing redshift.  To account for this and
construct an unbiased estimate of the projected satellite number
density profile over all primary galaxies, we accumulate the area
contributed by the $i$th primary to the $j$th annulus for the
detection of satellites brighter than $M_{\rm sat}^{\rm trun}$,
  \begin{equation}
     A_{ij}^{\rm p}(M_{\rm sat}^{\rm trun})=\left\{
  \begin{array}{ll}
      A_{ij} &  M_{\rm sat}^{\rm trun} < M_i^{\rm{lim}} \\
      0 &   M_{\rm sat}^{\rm trun} > M_i^{\rm{lim}} \ .
  \end{array}\right.
  \end{equation}
Here $A_{ij}$ is the unmasked area of the $j$th annulus surrounding the
$i$th primary and $M_i^{\rm lim}$ is the absolute magnitude that
corresponds to the apparent magnitude limit, $m^{\rm{lim}}$, of the SDSS
photometric catalogue at the redshift of the primary, $M_i^{\rm
lim}=m^{\rm{lim}}-5\log_{10}(D^L_i)-K(z_i)$, where $D^L_i$ and
$K(z_i)$ are the corresponding luminosity distance and $k$-correction.
This contributing area is set to zero if any potential satellites within the
magnitude bin are too faint to be included, in which case we exclude
this primary and its satellites as its contribution to the mean
projected satellite number density profile would be incomplete.  We
further define $N^{\rm sat}_{ij}(M_{\rm sat}^{\rm trun})$ to be the
number of detected potential satellites brighter than $M_{\rm
sat}^{\rm trun}$ in the $j$th annulus surrounding the $i$th primary
and $N^{\rm bck}_{i}(M_{\rm sat}^{\rm trun})$ to be the corresponding
number of detected galaxies in the outer annulus, $\ri<r<\ro$, whose
unmasked area is $A^{\rm outer}_i$. Hence we can express the mean
surface density of satellite galaxies brighter than $M_{\rm sat}^{\rm
trun}$ in the $j$th annulus as
  \begin{equation}
  {\emlsat}=\frac{\sum_i N^{\rm sat}_{ij}(M_{\rm sat}^{\rm
      trun})}{\sum_i A_{ij}} 
 - \frac{\sum_i N^{\rm bck}_i}{\sum_i A^{\rm outer}_i  } .
  \label{eqn:est}
  \end{equation} 

  In practice, we divide the projected radial distance from the
  primary into $20$ bins ($j=1,2,\cdots,20$).  Because of a 
  concern that the SDSS data reduction pipeline may occasionally
  misclassifies fragments of the spiral arms of bright galaxies as
  separate galaxies we exclude individual annuli that are within 1.5
  times the Petrosian radius, $R_{90}$, of the primary galaxy.  We set our
  magnitude limit, $\msatt$, either by absolute value, such as $-20,
  -19, -18$, or by magnitude relative to the corresponding primary,
  $\msatt=M_{\rm p}+1.0, 2.0, 3.0$. We present results using both
  thresholds so that we can determine whether the number density
  profiles depend on the absolute luminosity of satellites or on the relative
  luminosity between satellites and their primaries.

  The process of estimating the projected satellite number density profiles 
  is quite similar to estimating the satellite luminosity 
  functions in Paper~I. We divide our primaries into 
  three luminosity bins centred on $\satmc=-21.25, -22.0, -23.0$. The choice
  of the parameters $\ri$ and $\ro$ are a balance between making
  them sufficiently large to avoid severely truncating the density
  profiles and making them too large such that our sample shrinks
  due to the selection process excluding overlapping systems.
  A sensible choice is to set $\ri$ to exceed the anticipated size of the satellite system, $r_{200}$\footnote{Here $r_{200}$ depicts the radius at which the mean interior density is $200$ times the cosmological critical density.},
  and $\ro$ to be roughly a factor of two larger so as to get a good, but
  still local, estimate of the background density. Here 
  we have adopted the following
  values of $(\ri,\ro)$,  $(0.3,0.6),(0.4,0.8)$ and
  $(0.55,0.9)~{\rm Mpc}$ for primaries in magnitude bins, $\satmc=-21.25,-22.0$ and
  $-23.0$, respectively. The values of $\ri$ have been compared
  with the mean of the estimated $r_{200}$ values for each galaxy in
  the chosen magnitude bin (see Section 3) and are found all to be
  larger, suggesting that the search radii for the different primary
  magnitude bins are sufficiently large to capture all satellites.
  Additional reassurance is provided by the tests
  in Appendix~\ref{appendix:c}, which show that the profiles are insensitive
  to changes in the values of $\ri$ or $\ro$.

%    Therefore the parameters that we will use 
%    in the rest of the paper are, \\
%    {\small
%    \begin{tabular}{p{8mm}p{6mm}p{6mm}p{6mm}p{8mm}p{4mm}p{4mm}p{7mm}}
%    $\satmc$ & $\ri$ & $\ro$ & $\satdmbin$ & $\satdmf$ & 
%    $\satdzs$ & $\sGatap$ & $m_{\rm v}^{\rm lim}$\\
%    -21.25 & 0.3 & 0.6 & 0.5 & 0.5 &0.002&2.5&-20.5\\
%    -22.0 &  0.4 & 0.8  & 0.5 & 0.5 &0.002&2.5&-20.5\\
%    -23.0 &  0.55 & 0.9 & 0.5 & 0.5 &0.002&2.5&-20.5\\ 
%    \end{tabular}}.\\

\subsection{Exploring the angular distribution of satellites}

\begin{figure}
	    \begin{center}
	     \includegraphics[width=70mm]{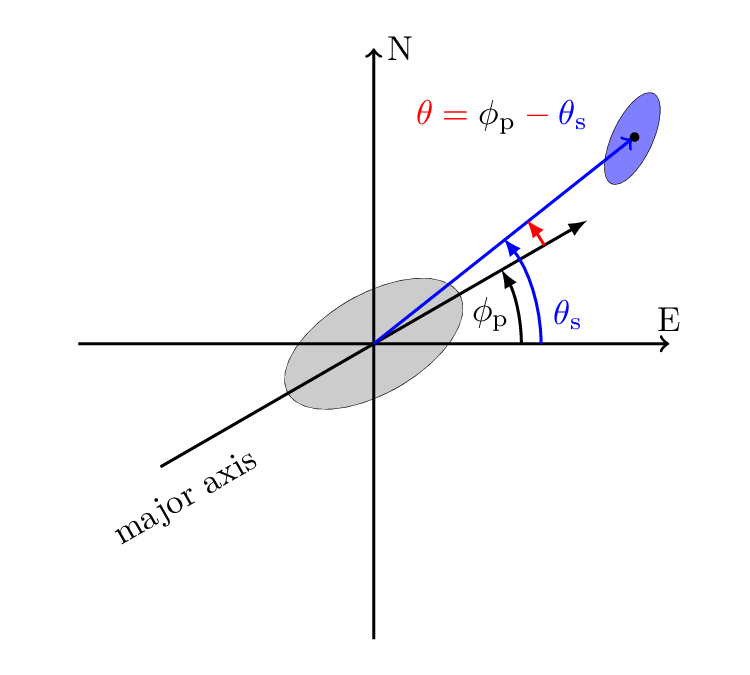}
	     \caption{Schematic showing the definition of the position angle
	     $\theta$ which characterises the angular position of
	     the satellite relative to the direction of 
	     the primary galaxy's major axis. The grey and blue ellipses are
	     the primary and satellite galaxy, respectively.
	     The angle $\phi_{\rm p}$ is the position angle of the primary.
	      }
	     \label{fig:theta}
	 \end{center}
	\end{figure}

The projected radial satellite number density profile, $\Sigma^{\rm
sat}(r)$, which is the focus of this paper, is the azimuthal average
of the 2D surface density, $\Sigma^{\rm sat}(r,\theta)$, where $\theta$
can be taken as the position angle between the major axis of the primary
and the line connecting the primary and satellite (see
Fig~\ref{fig:theta}).  The angular dependence of this distribution may
also carry information on the formation and evolution of the satellites
around their primaries.  For example, if we assume that the satellite
galaxies inhabit an unbiased set of dark matter subhalos, then we would
expect satellites to cluster preferentially along the major axis of
the halo \citep{libeskind05}.  Moreover, it is known that the host halos of satellite
systems are accreted from filaments, which can cause the angular
distribution of satellites around primaries to be anisotropic
\citep{har00}.  In fact, numerous such anisotropies have been
observed.  For example, the famous ``Holmberg Effect'' \citep{hol69},
suggests satellites of isolated, large, and inclined spiral galaxies
are preferentially located along the minor axes of their primaries, a
result supported by \citet{zar97b}.  However \citet{yan06},
\citet{azz07}, \citet{bra05} and \citet{agu11} found the opposite
effect that satellites prefer alignment with the major axis,
especially for the satellites of red primaries. Since the direct
observation of satellite systems is not easy, the sample of external
satellite systems is limited in both volume and quality.  These
contradictory results may suggest that the mean amplitude of the
anisotropy could be very weak, or the form of the anisotropy could be
dependent on the selection of the primaries or even the satellites
themselves.

It is therefore interesting to attempt to quantify the mean anisotropy
of our large sample of satellite galaxies.  We characterise the
angular distribution using the position angle $\theta$ described in
Fig.~\ref{fig:theta}.  The assumed elliptical symmetry of the primary
implies that $\theta$ ranges from $0^\circ $ to $90^\circ$ with these
extremes indicating that the satellites are located along the major or
minor axis respectively. The anisotropy of the angular distribution is
then quantified by the probability distribution of the angle
$\theta$. In practice, accurate measurement of $\theta$ requires a
robust measurement of the position angle $\phi_{\rm p}$ defining the
orientation of the primaries. To achieve this, we adopt the same
selection criteria as \citet{siv09}. We only select primaries and
their satellites that satisfy the condition, $q_{\rm iso}<0.9$ and
$q_{\rm mom}<0.9$, where $q_{\rm iso}$ is the isophotal axis ratio defined
as $q_{\rm iso}=a_{\rm iso}/b_{\rm iso}$ and $q_{\rm mom}$ is the
adaptive moments axis ratio, $q_{\rm mom}=((1-e)/(1+e))^{1/2}$ where
$e=(e^2_++e^2_{\times})^{1/2}$ \citep{ryd04}\footnote{$e_+$ and $e_{\times}$
are second-order parameters from SDSS, where $\tau=M_{xx}+M_{yy},
e_{+}=(M_{xx}-M_{yy})/\tau$, $e_{\times}=2M_{xy}/\tau$ and $M_{XX}$
here are the second-order adaptive moments.}.  We also exclude the
primaries together with their satellites if there is a discrepancy of
more than $15$~degrees between the measured isophotal and de
Vaucouleurs position angles, $\Delta \theta_{\rm p}^{\rm
  iso-mod}>15^\circ$. 

For our sample of selected satellite systems, the number of satellites
located at angle $\theta_j$ around the $i$th primary, can be estimated
as
\begin{equation}
  N^{\rm sat}_i(\theta_j)=N_i^{\rm inner}(\theta_j)-
  \frac{\ai}{\aio} \left\langle N_i^{\rm outer}(\theta_j)\right\rangle,
\end{equation}
where $\left\langle N_i^{\rm outer}(\theta_j)\right\rangle$ represents
the azimuthal average as we assume the background galaxies are, 
on average, isotropically distributed. Note that to avoid biasing the
angular distribution we exclude systems that are incomplete due to
the survey mask.
We can then define an    
unbiased estimator of the average distribution
of the satellite position angle $\theta$ for all selected primaries as
\begin{equation}
	 \widetilde N(\theta_j,\msatt)=\frac{\sum_iN^{\rm sat}(\theta_j,\msatt)}
	 {N_j^{\rm prim}(\msatt)} .
\end{equation}

   \begin{figure}
	    \begin{center}
	    \includegraphics[width=84mm]{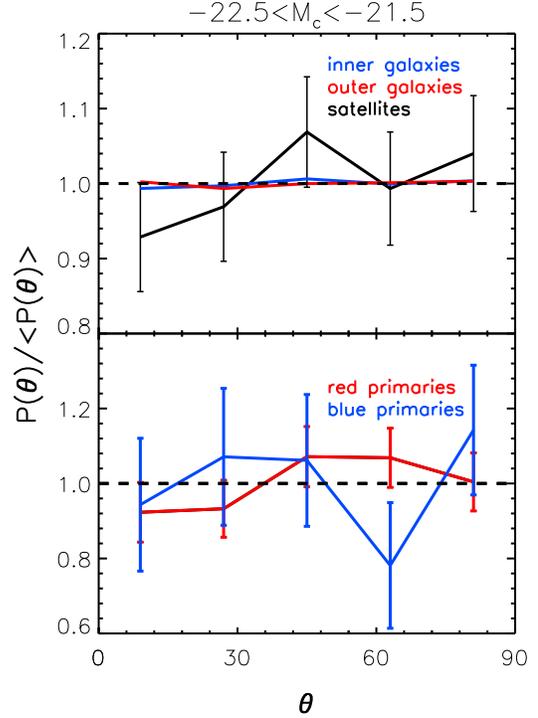}
	    \caption{The probability distribution, $P(\theta)$, of the position
	    angle $\theta$ of satellite galaxies. In the top panel the
          blue and red lines show the distributions, $P^{\rm
            inner}(\theta)$, of inner galaxies, and $P^{\rm
            outer}(\theta)$, of outer galaxies, respectively. The inferred
          distribution for true satellite galaxies, 
          $\widetilde P^{\rm sat}(\theta)$, is shown by the black line.
           The lower panel shows the distributions
	    for red and blue primary subsamples, as the red and blue
            solid lines respectively. 
	    The error bars are calculated from $1000$ bootstrap resamplings.
	    The expectation for a uniform distribution 
          is shown by the dashed lines in the two panels.
	     }
	    \label{fig:pd}
	    \end{center}
	\end{figure}

\begin{figure*}
   \includegraphics[width=142mm]{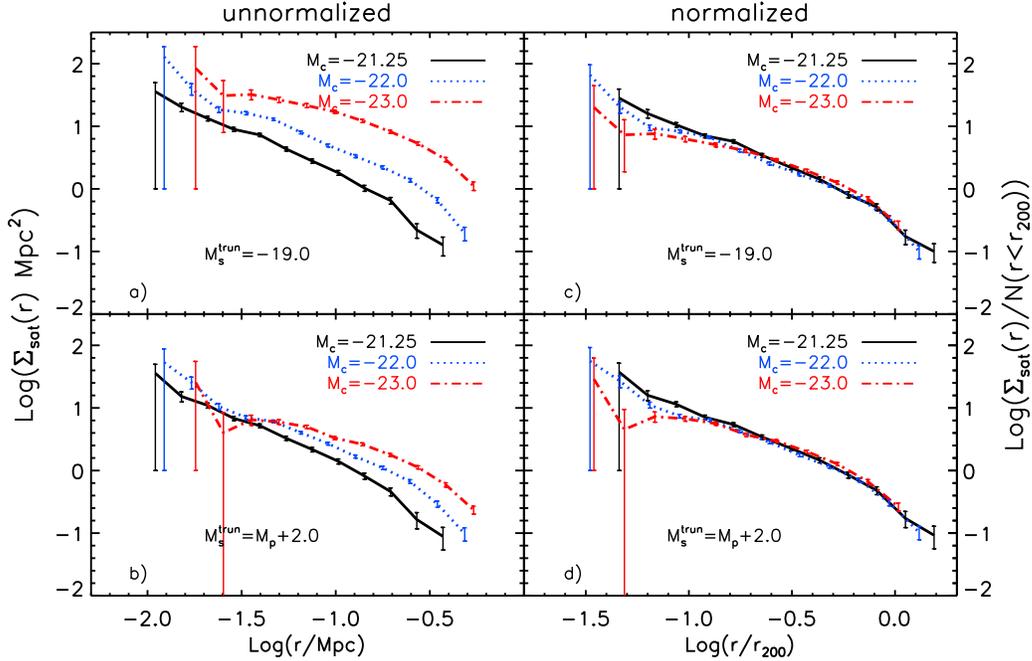}
   \caption{The mean projected number density profiles of satellites
     around primaries for various selections of primaries and
     satellites. The left hand panels (a and b) show the mean number
     densities and the right hand panels (c and d) show the same profiles
     but now normalized by the total number of satellites within the
     $r_{200}$ radius (see right-hand axis) and with the radius
     expressed in units of the adopted mean $r_{200}$ radius of the primaries.
     In each panel the different
     coloured lines correspond to primaries of differing luminosity
     as indicated in the legend.  The profiles in the upper panels (a
     and c) are for satellite samples brighter than $\msatt= -19$,
     while the lower panels (b and d) are for satellites that are
     less than 2.0 magnitudes fainter than their respective
     primaries. The error bars show the errors on the mean profiles
     estimated by bootstrap resampling.}  
   \label{fig:fig_nor}
\end{figure*}

The normalized probability distribution of $\theta$ for primaries in the
magnitude bin $\satmc=-22.0$ is shown in Fig.~\ref{fig:pd}.  The blue,
red and black solid lines in the top panel are the probability
distributions of $\theta$ for inner galaxies, $P^{\rm inner}(\theta)$,
outer galaxies, $P^{\rm outer}(\theta)$, and that inferred for
satellite galaxies, $\widetilde P(\theta)$, respectively. In the
bottom panel of Fig.~\ref{fig:pd}, we show the probability
distribution of $\theta$ for two satellite subsamples split by rest
frame colour of their primary galaxies. These samples are divided
according to the well-known colour bimodality in the colour-magnitude
plane \citep[e.g.][]{str01,bal04,zeh05}.  Following \cite{zeh05}, we
use an equivalent colour criterion of
$^{0.0}(g-r)_{\rm{cut}}=0.19-0.24M_r$ (not identical to
\citeauthor{zeh05} as our magnitudes are $k$-corrected to $z=0.0$
rather than $z=0.1$).  The probability distributions of $\theta$ in
Fig.~\ref{fig:pd} are all consistent with isotropic distributions.
This is confirmed by a two-sample Kolmogorov-Smirnov (KS) test that
compares the distributions of inner and outer galaxies for the whole
primary sample and also the subsamples of red and blue primaries.  The
two-sample KS probabilities from the whole sample, blue and red primary
subsamples are $0.64, 0.37, 0.87$ respectively, which implies that the
pairs of distributions have no statistically significant differences.
The same tests for primaries in other magnitude bins show
similar results.

Therefore, with  our satellite system sample, we find that there is
no statistically significant evidence that the distribution of
satellites around primaries is anisotropic.  
This could signify that the anisotropy of the distribution of satellites around
isolated primaries is intrinsically insignificant.  However, one also has
to keep in mind that our inner sample includes contamination from
interlopers, because these are only rejected using inaccurate photometric
redshifts, and this could dilute any intrinsic anisotropy signal.

\section{Results}
\label{}

We now return to the azimuthally averaged density profiles.  
Density profiles for satellites brighter than $\msatt$ around
primaries of magnitude $M_{\rm c}$ are shown in Fig.~\ref{fig:fig_nor}
for a variety of primary magnitude bins and satellite magnitude cuts.
Panels~\ref{fig:fig_nor}a and \ref{fig:fig_nor}b show that the number
of satellites increases with increasing primary luminosity and extends
to larger radii. 
To investigate the variation in profile shape between 
different subsamples of satellites and primaries, it is helpful to
use scaled variables. To this end, we recast the profiles in
terms of $x=r/r_{200}$ and divide the number densities by the total
number of satellites within $r_{200}$. 

The values of $r_{200}$ used to scale the radii can be determined from
the stellar masses, themselves inferred from the measured galaxy
luminosities and colours, and the abundance matching technique of
\citet{guo10}, which gives
\begin{equation}
    \frac{M_*}{M_{\rm halo}}=c \left[\left(\frac{M_{\rm halo}}{M_0}\right)^{-\alpha}+
    \left(\frac{M_{\rm halo}}{M_0}\right)^{\beta} \right]^{-\gamma},
\end{equation}
where $c=0.129$, $M_0=10^{11.4}~\rm{M_{\sun}}$, $\alpha=0.926$, $\beta=0.261$ and 
$\gamma=2.440$ are fitted constants.
The halo mass can be related to a radius through
\begin{equation}
    M_{\rm halo}=\frac{4\pi}{3}200\rho_{\rm crit}r^3_{\rm 200}.
    \label{eqn:vir}
\end{equation}

\begin{figure*}
 \includegraphics[width=152mm]{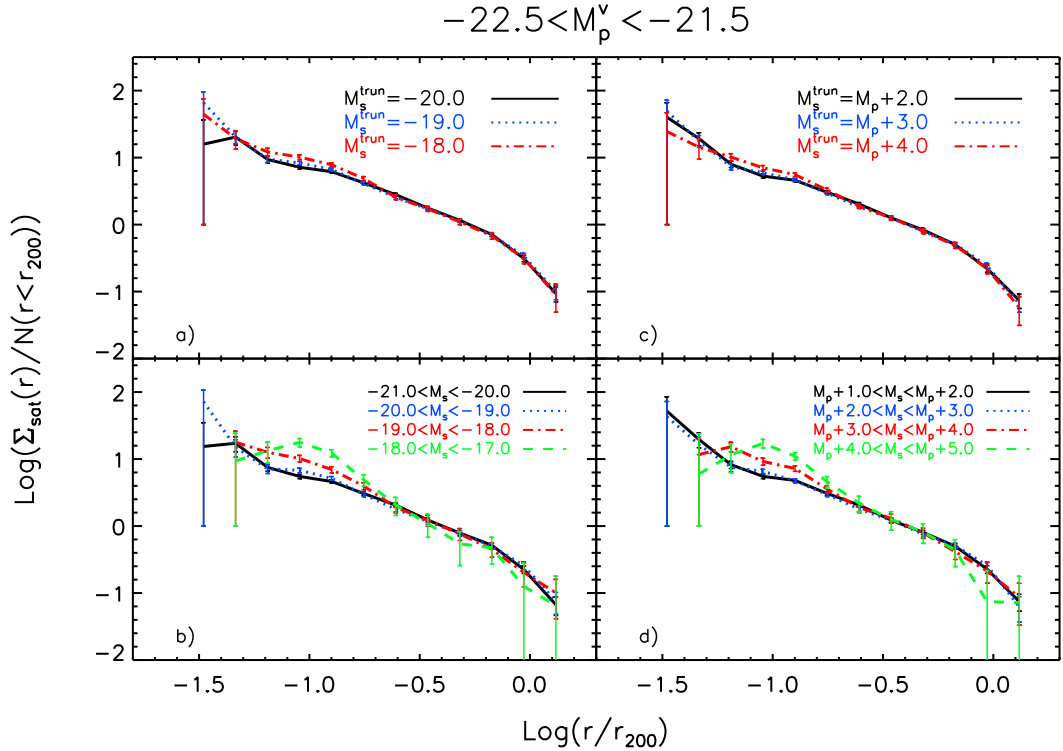}
 \caption{The dependence of the scaled satellite density
          profiles on satellite luminosity for primaries in the
          magnitude range $-22.5<M_{\rm p}<-21.5$.  The different
          panels show different satellite selections. The upper
          panels, which are very similar, show profiles for satellites
          brighter than a threshold that is either a fixed value (panel~a)
          or specified as  a magnitude difference with respect to
          the corresponding primary (panel c; see legend). The lower two
          panels show profiles for satellites in bands of magnitude
          again either specified between fixed values (panel b)
          or between values relative to the corresponding primary (panel d).
          }
 \label{fig:pro_same_p}
 \end{figure*}

 For the primary galaxies there is a significant uncertainty in the
 stellar mass that is inferred from the measured luminosities and
 colours. Thus, rather than using the individual $r_{200}$ values to
 normalise satellite number density profiles for each primary before
 stacking them together, the mean $r_{200}$ for all primaries in the
 luminosity bin of interest is determined and a single rescaling is
 performed on the unscaled, stacked profile. Using $r_{200}$ values
 determined in this way, the $M_c=-21.25$ and $-22.0$ samples line up
 very well, as shown in panels~\ref{fig:fig_nor}c and
 \ref{fig:fig_nor}d. However, the $M_c=-23.0$ results, not shown in
Fig.~\ref{fig:fig_nor}, are slightly offset.
For these, the most luminous primaries, the relation between stellar mass
and halo mass becomes very flat and there is a large spread in the halo
mass corresponding to a given stellar mass. This makes the assignment of a
value of $r_{200}$ to these primaries extremely uncertain. The directly
inferred virial radius is 0.73~Mpc but we find that a smaller value of
0.52~Mpc results in better scalings. Given the large uncertainty in this
assignment, it is not unreasonable to adopt this smaller value. We shall
do this in what follows but this uncertainty must be borne in mind when
interpreting the results for the brightest primary bin.
%For these, the most luminous
 %primaries, there is a large spread in halo masses at a given stellar
 %mass, so one might expect a greater uncertainty in the $r_{200}$
 %value for this sample. Thus, we empirically choose $r_{200}$ so that
 %the satellite number density profile lines up with those of the lower
 %luminosity primaries, as shown in the right hand column of
 %Fig.~\ref{fig:fig_nor}. 
 The final values of $r_{200}$ for the three
 primary magnitude bins are $0.24,0.37,0.52~{\rm Mpc}$, the first two
 of which come directly from equation~(\ref{eqn:vir}). By adopting
 these mean $r_{200}$ values, we have a mass-to-light ratio increasing
 with luminosity as $M/L_{V}\propto L_{V}^{0.42}$. This is similar to
 the relation found by \cite{pra03}, albeit for the B-band, from a set
 of spectroscopically selected satellites from SDSS. For $M_{V}$ =-22,
 we find that $M/L_{V}$=140.

First, we explore the dependence of the normalized profiles on the
luminosity of the satellites.  In Fig.~\ref{fig:pro_same_p}, we show
normalized profiles for primaries of fixed luminosity
($-22.5<M_{\rm p}<-21.5$) with a variety of different satellite
selections. In all cases we find the outer shapes of the density
profiles to be very similar. The only variation is on small scales
(roughly $r/r_{200} < 0.1$) where the density profile is steeper
and higher for the faintest satellites. 

\begin{figure*}
 \includegraphics[width=152mm]{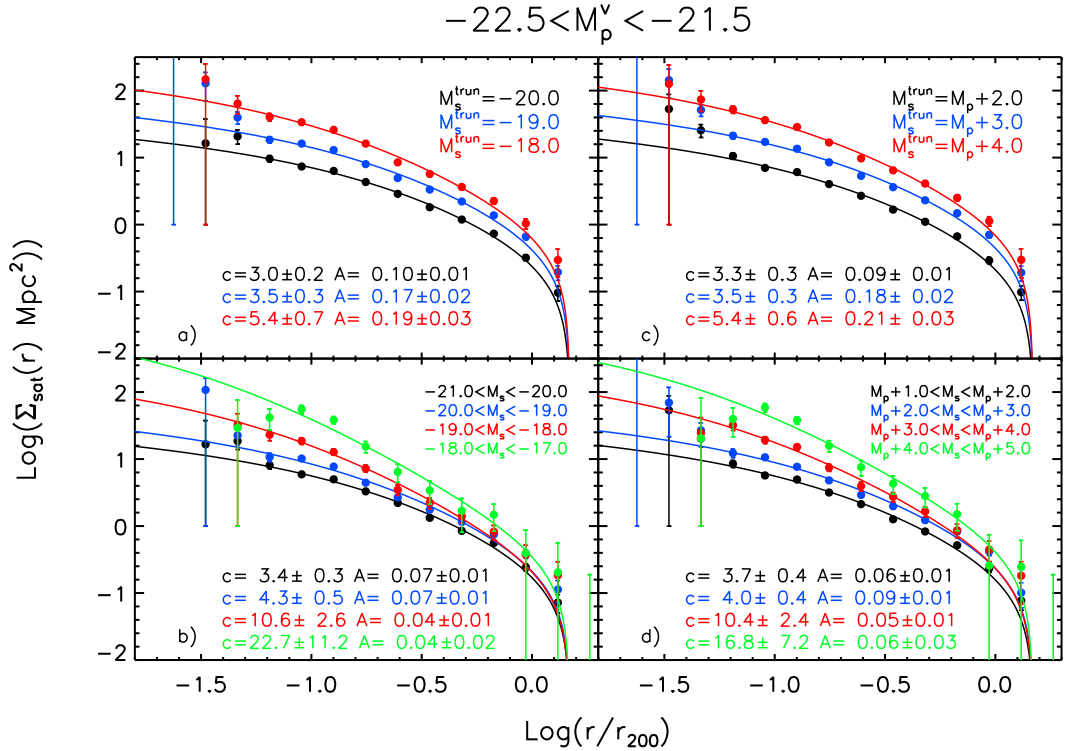}
 \caption{Fits to the satellite density profiles of primaries of
   magnitude $-22.5<M_{\rm p}<-21.5$ using projected,
   background-subtracted NFW profiles. The various panels show
   different selections of satellites as in
   Fig.~\ref{fig:pro_same_p}, except that here we
   have not normalized the profiles, but instead allowed the amplitude,
   $A$, of the fitted profiles to float. The measured profiles
   are shown by the data points and the best fitting NFW profiles are 
   plotted as solid lines. The best fitting amplitudes
   and concentrations are listed in the legends in each panel.
}
  \label{fig:pro_same_p_fit}
\end{figure*}

We examine the shape of these density profiles more systematically in
Fig.~\ref{fig:pro_same_p_fit} where we fit the unnormalized density
profiles using an analytic model.  We have chosen to fit our satellite
profiles using NFW profiles as they are known to be good fits to both
dark matter haloes \citep{nfw96,nfw97} and to the distribution of
substructures within them \citep{die04,spr08,lud09}. To perform these
fits we first project the NFW profile and subtract from it the mean
density in an outer ``background'' annulus as described in Appendix
\ref{appendix:a}, so as to match the background subtraction process
that we have applied to our observational data. We then perform a fit
using $\Sigma^{\rm sat}(r)=A\hat\Sigma(r,c,r_{\rm 200})/M_{\rm 200}$,
where $A$ is a scale factor, $c$ is the concentration, and
$\hat\Sigma(r,c,r_{\rm 200})$ is the projected NFW profile with
background subtracted as given in Eqn.~\ref{eqn:nfw}.  The
$r_{200}$ radius is fixed to the respective values that we
have adopted for each of our bins of primary magnitude.

Fig.~\ref{fig:pro_same_p_fit} shows the resulting fits.  The bright
satellites, such as satellites with magnitudes $-21.0<M_s<-20.0$ or
$M_{\rm p}+1.0 <M_s<M_{\rm p}+2.0$ are very well fitted by the
projected and background subtracted NFW profiles.  The NFW fits also
remain good descriptions of the data for the cumulative samples of
satellites defined by a faint magnitude threshold. For these samples, 
shown in the upper panels of Fig.~\ref{fig:pro_same_p_fit}, the
concentration increases steadily with decreasing luminosity. In the
lower panels of Fig.~\ref{fig:pro_same_p_fit}, which show density
profiles for satellites in differential bands of luminosity, we see
both a stronger dependence of concentration on luminosity and small
deviations from the NFW form for the faintest satellite samples.

\begin{figure*}
 \includegraphics[width=148mm]{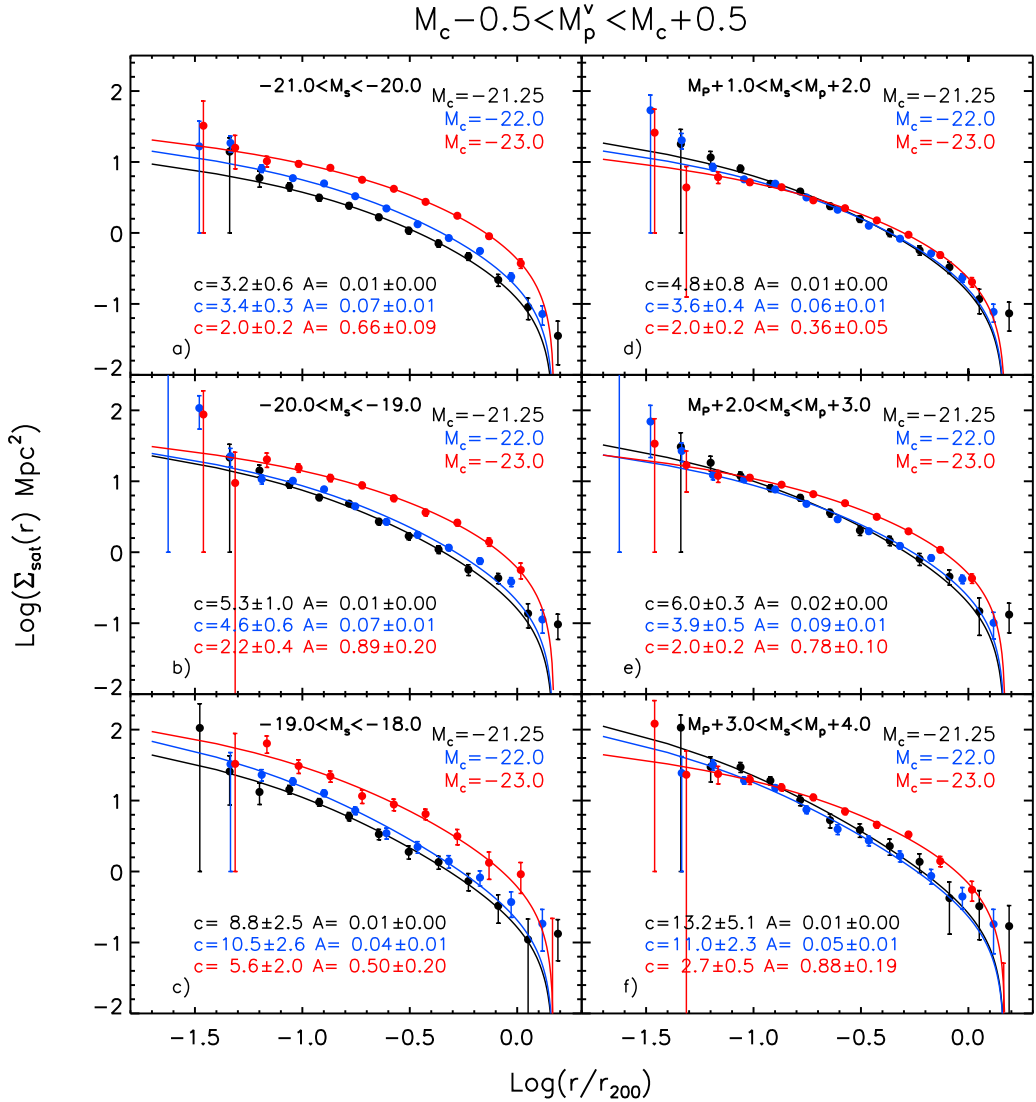}
 \caption{ NFW fits to the density profiles of satellites around
   primaries of different luminosity.  Each panel shows fits for the
   three different bins of primary luminosity indicated in the
   legends.  The different panels correspond to different selections
   of satellites.  The panels on the left, a), b) and~c), are for
   successively fainter bands of satellite luminosity as indicated on
   the legend, with the range of satellite magnitudes being the same
   for each primary. In contrast the panels on the right, d), e)
   and~f), are for bins of satellite magnitude that are specified as
   an offset relative to the magnitude of their respective primary.  }
 \label{fig:pro_diff_p_fit}
\end{figure*}

We now turn to the dependence of the satellite profiles on the
magnitude of the primaries. In Fig.~\ref{fig:pro_diff_p_fit} we show
fits of the projected and background-subtracted NFW profiles to
satellite profiles of primaries in each of our three magnitude
bins. Each of the panels corresponds to a different selection of
satellites.  We see that NFW fits are good descriptions of the
satellite distribution regardless of the luminosity of the primary.
The right hand panels of Fig.~\ref{fig:pro_diff_p_fit} show the
density profiles and fits for sets of satellites defined by fixed
offsets in magnitude from the magnitude of their respective
primary. If the combined primary and satellite systems scaled in a
self-similar way we would expect the three density profiles in each of
these panels to lie on top of each other. In contrast, in each panel,
we see systematic variations in the shape and amplitude of the
profiles with the primary luminosity. If instead we look at the left
hand panels, which show satellites selected in different fixed
magnitude bands, then we see that the concentration decreases steeply
with increasing satellite luminosity, but is less dependent on the
luminosity of the primary. With the exception of the brightest primary
magnitude bin ($\satmc=-23.0$), the satellites of a given luminosity
are more or less distributed in the same way about primaries of
different luminosity. Only the normalization of the profile, as
parameterized by $A$, increases with increasing primary
luminosity. This depends quite strongly on luminosity going roughly as
the luminosity to the power of $2.5$.  If we normalized each of these
satellite profiles as we did in Fig.~\ref{fig:pro_same_p}, then their
shapes would show very little variation with primary luminosity.

For the case of primaries in the $\satmc=-23.0$ bin, the measured
concentrations for the satellite distributions are 
systematically lower than those of similar luminosity satellites
around less luminous primaries. The reason for this is not clear. It
may be that this result reflects the actual satellite distribution
around bright primaries. However, it could be an artefact of the
estimation procedure. One possibility is that the non-linearity of the
halo mass-stellar mass relation \citep{guo10} means that the range
of actual halo mass increases in the brightest primary bin. Thus,
stacking all primaries using a single $r_{200}$ value may be introducing
errors that would smear out the resulting profile.
%makes our
%estimation of the halo masses and hence $r_{200}$ radii of the most
%massive primaries particularly uncertain, leading to errors in the
%inferred number density profile. 
Another potential source
of systematic error comes from the tendency for faint satellites to be
missed around bright primaries because of inaccuracies in the
sky-subtraction \citep{man05}. Even with the updated sky-subtraction
algorithm employed for DR8 there may still be some residual loss
of faint satellites around the brightest primary galaxies \citep{aih11}.

\subsection{Colour and type dependence}

\begin{figure*}
\includegraphics[width=148mm]{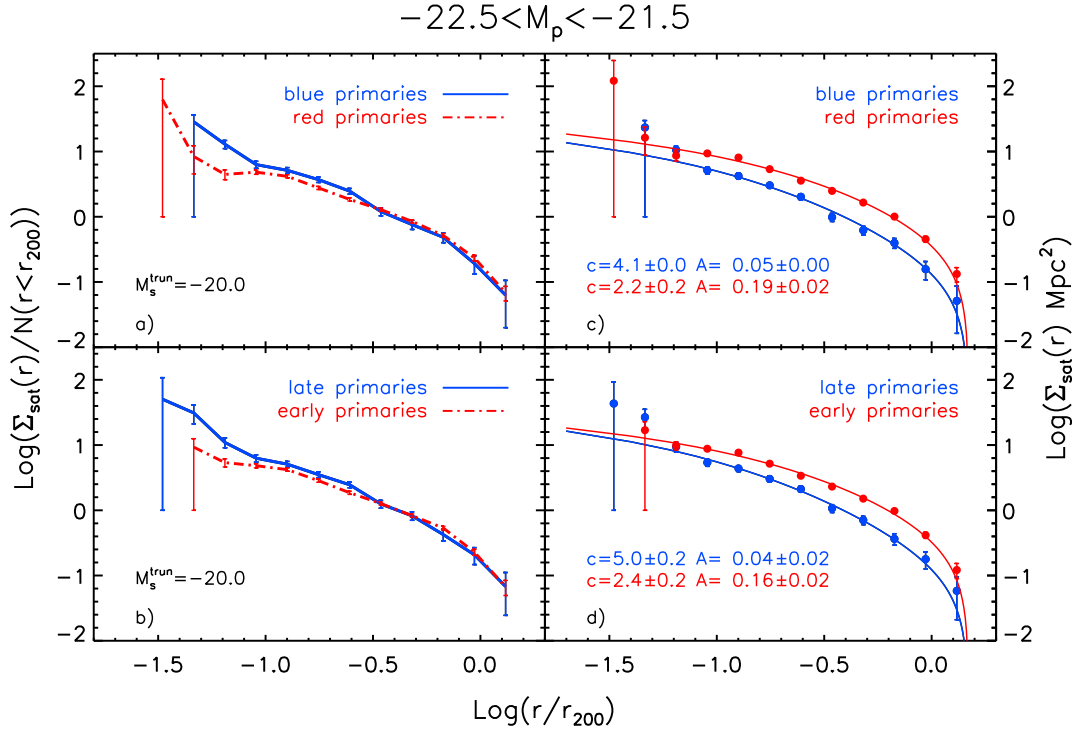}
\caption{The satellite profiles for primary galaxies
           of magnitude $-22.5<M_{\rm p}<-21.5$
           split by the type
         (concentration) and colour of the primary. 
         Panels a) and~b) show the normalized profiles
         while panels~c) and~d) show NFW fits to the
         unnormalized profiles (see right hand axis).
         In the upper panels the blue lines refer to blue primaries
         and red lines to red primaries, while in the lower panels
         blue refers to late-type primaries and red to early types. }
  \label{fig:pro_prim_sub}
\end{figure*}

Our large sample of satellite systems enables us to divide our samples
by the colour or the type of the primaries.
Fig.~\ref{fig:pro_prim_sub} shows the resulting profiles when
primaries of $V$-band magnitude $-22.0\pm 0.5$ are split by colour and
by concentration. Panels a) and~c) of Fig.~\ref{fig:pro_prim_sub} show
that the normalized profile of satellites around blue primaries is 
more concentrated than that around red primaries. Panels b) and~d) 
split the sample into early and late types, where early type is
defined as having a concentration index
$\mathcal{C}\equiv{petroR_{90}}/{petroR_{50}}\ge 2.6$, with
$petroR_{90}$ and $petroR_{50}$ being the SDSS Petrosian 90\% and 50\%
light radii respectively. This division roughly separates early type
(E/S0)  from late-type (Sa/b/c, Irr) galaxies
\citep{shi01}. We also see the amplitude of the profiles of late types
is suppressed with respect to that of the early types. However, the
concentration indices, $c$, from the fits similarly show that the concentration of
satellites around late types is higher than that of early types.

\begin{figure*}
   \includegraphics[width=138mm]{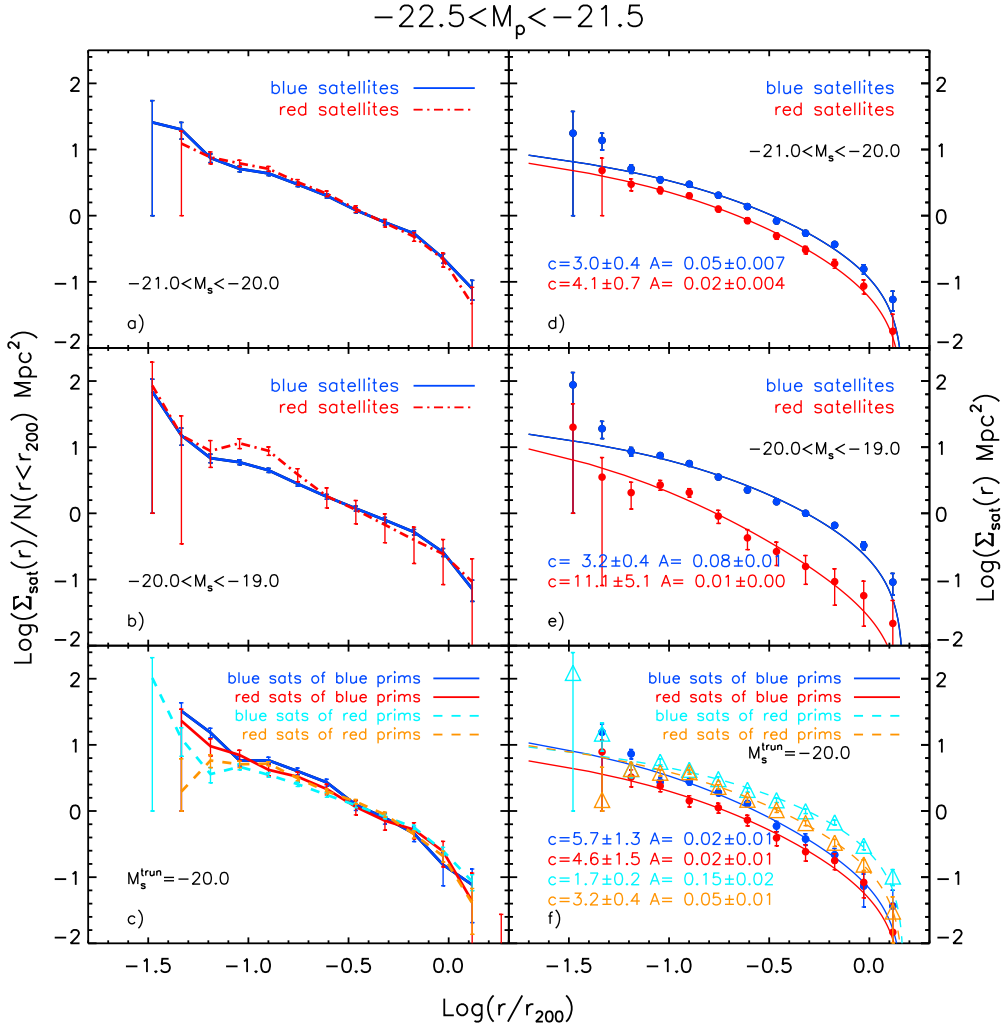}
   \caption{Satellite density profiles split by the colour and
     luminosity of the satellites. Panels a) and b) compare the 
     profiles of blue (blue solid line) and red (red dot-dashed line)
     satellites for two different bins of satellite luminosity (see
     legend). Panels d) and e) show the corresponding unscaled
     profiles as red and blue symbols together with curves depicting 
     NFW fits. The best fitting amplitudes and concentrations are
     given in the legends. For satellites brighter than -20.0, 
      panels c) and f) compare the profiles 
     of red and blue satellites around red and blue primaries.
     The profiles for blue primaries are shown with solid
     lines and those for red primaries with dashed lines. The 
     blue and cyan lines are for blue satellites and red and orange
     lines for red satellites as detailed in the legends.
     In panel f) the smooth curves show the NFW fits to the
     corresponding unscaled profiles
     and again the
     best fit parameters are listed in the legend.
}
   \label{fig:pro_sat_sub}
\end{figure*}

We can also use the colour information available in SDSS to probe the
properties of the satellites. Firstly, for the bin of primary
magnitude, $\satmc=-22.0$, we divide the satellites into two luminosity
bins, $-21.0<M_{\rm s}<-20.0$ and $-20.0<M_{\rm s}<-19$ and into red
and blue subsamples using the same cut as before.
Fig.~\ref{fig:pro_sat_sub}a,b,d and~e show the measured profiles of these
blue and red satellites and NFW fits.  We first note from
Fig.~\ref{fig:pro_sat_sub}d and~e that for these relatively bright
satellite samples the abundance of blue satellites is greater than
that of red satellites at all radii, with this difference increasing
for the fainter sample. The profiles of the brighter satellites have a
similar shape for red and blue satellites, while the fainter red
satellites have an excess at $\sim 0.1r_{200}$ relative to the fainter
blue satellites.
%Beyond this we see that
%profiles of the slightly fainter satellites ($-20.0<M_{\rm s}<-19$)
% and their fitted concentrations are significantly different.  The red satellites have  more
%concentrated profiles  at $r\lsim 0.1 r_{\rm
% 200}$ relative to blue satellites and so are preferentially concentrated close to the
%primary. In contrast the blue primaries have a lower concentration
%with a 
%more power-law like extended distribution.  
To investigate whether
these differences are driven by the colours of the associated primary
galaxies we further split the satellites brighter than
$M_{\rm s}^{\rm trun}< -20$ by the colour of their primary. The results
are shown in Fig.~\ref{fig:pro_sat_sub}c and~f. Both red and blue
primaries have more blue than red satellites. The concentrations of
red and blue satellites around blue primaries are similar. Red
primaries have lower concentrations for both their red and blue
satellites, with the blue satellites having a particularly low
concentration. The 
colour of the primary appears to be more important than that of the
satellite in determining the concentration of the satellites. As shown
in Fig.~\ref{fig:pro_diff_p_fit}, the satellite luminosity also has a
strong effect. 

%It suggests that
%the abundance of blue satellites around any color of primaries are 
%both greater than that of blue satellites. However, the blue and 
%red satellites more or less have similar concentrations, while the
%blue satellites around red primaries are less concentrated than the 
%red satellites around red primaries.
%the properties of the satellites themselves rather than the properties
%of the primaries. We do note, as in
%Paper~I, that generally the abundance of satellites is greatest around
%red primaries.

\section{Discussion}

Using a stacking analysis we have estimated the mean projected density
profiles of satellite galaxies around a large sample of isolated
primary galaxies selected from the SDSS DR8 spectroscopic galaxy
catalogue and we have quantified how they depend on the properties of the
satellites and primaries.  The selection of primaries and the local
background subtraction technique, which makes use of photometric
redshifts, is the same as in Paper~I \citep{gu11} where we estimated
the mean satellite luminosity functions of these systems.  Our main
conclusions are:

(i) We find no evidence for any anistropy in the satellite galaxy
  distribution relative to the major axes of the primaries.

(ii) The projected number density profiles of satellites
brighter than a $V$-band magnitude of $-17$ are well
determined for three separate bins of primary magnitude,
$-21.25, -22,0, -23.0$.

(iii) Apart from the faintest satellites, for which there is 
a slight excess at small galactocentric projected distance, all other
density profiles are well fitted by projected NFW profiles
 that have been background subtracted
to match the procedure that has been applied to the data.

(iv) The concentration of the NFW fits decreases systematically with
increasing satellite luminosity and is almost independent of the
luminosity of the primaries (see Fig.~\ref{fig:pro_diff_p_fit}).  
Thus, bright satellites have more extended
distributions and fainter satellites are more centrally concentrated.
%For the faintest satellites, $M_{\rm s}>-19$, a clear excess or bump
%above the NFW profile is seen at small galactocentric projected
%distances for all primary magnitude bins.

(v) The radial distribution of satellites is dependent on
the colour and morphology of their primaries. Satellites are more
numerous around red/early primaries and have more extended, lower
concentration, distributions (see Fig.~\ref{fig:pro_prim_sub}).

(vi) The radial distribution of satellites also depends on the
colour of the satellites. Blue satellites are more numerous than red
satellites at all radii (for the luminosity range we probe) and faint red
satellites are more centrally concentrated (higher NFW concentration)
than faint blue satellites. Further sub-divided samples show that the
concentration of the blue or red satellite profile depends more on
the colour of the primaries than it does on the colour of the satellites.

As a check of potential systematic effects in our results, we have
also performed the same analysis using the SDSS DR7 dataset.
Generally, the results based on DR7 are consistent with those from
DR8, although we do observe some differences in the distribution of
faint satellites. This is most likely due to less accurate
photometric reduction and sky-background subtraction for DR7
(see Appendix~\ref{appendix:b} for more details).

With the advantage of our large and carefully selected samples, we
have discovered a variety of interesting information about the
projected number density profiles, which it has not been possible to
quantify clearly in previous work. However, even with a very limited
sample, a pioneering study by \citet{lor94} found that the
distribution of satellites is dependent on the morphology of
primaries. They found that the number of satellites around early-type
primaries is greater than that about late-type primaries and that 
the concentration of the satellite distribution is higher around early 
type primaries. We confirm the greater abundance of satellites around 
early-type primaries,
but contrary to \citet{lor94} we find higher concentrations for
satellite systems around late-type primaries.
More recently, \citet{van05}, \citet{sal05} and \cite{chen06} studied the
projected number density profiles of satellites of isolated galaxies
using larger samples from the Two Degree Field Galaxy Redshift Survey
(2dFGRS) and SDSS.  Although \citet{van05} cautioned that the profiles
from 2dFGRS were incomplete because of incompleteness in close galaxy
pairs, their study revealed that satellites followed NFW
profiles. \citet{sal05} found that the profiles of satellites depart
from a power law at small galactocentric projected distance, and that
they are dependent on the colour of the primaries, which is similar to
our conclusions (iv) and (v).  They also found the distribution of
satellites to depend on their colour, but argued that this may be caused
by the correlation between satellites and primaries. In our study,
conclusion (vi) shows that the distribution of satellites not only
depends on the properties of satellites, but also depends on the
colour of primaries.  \citet{chen06} and \citet{tol11} selected
samples only from the SDSS spectroscopic catalogue in their
studies. \citet{tol11} found the 3D number density profiles of
satellites can be fitted by a power-law with a slope $\rho \propto
r^{-1.8}$. After projecting, the slope of this density profile will be
close to ours. With a careful treatment of interlopers, they fitted
the profiles with a power-law form and found them to be independent of
the luminosity of the primaries. These conclusions are consistent with
ours.  Very recently, \citet{lar11} estimated the radial density
profiles of satellites around primary samples brighter than $-20.5$
and $-21.5$. They also found the amplitudes of the profiles depend on
the luminosity of the primaries, and the shapes on their colour.

The physics of the projected number density profiles of satellites
involves both the physics of the hierarchical assembly of dark matter
halos and the physics of the galaxy formation that occurs in these
assembling halos. Hence quantifying these profiles will help constrain
both galaxy formation models and the nature of the dark matter.  
We expect that our profile results and those of others will be an important
input into refining theoretical models and the next incarnation of
full N-body/gasdynamic simulation that can resolve the physics of the
formation of satellite galaxies.

\section*{Acknowledgements}

We thank Peder Norberg for supplying the mask and software for
quantifying the sky coverage of the SDSS DR7. 
QG acknowledges a fellowship from the European Commission's Framework
Programme 7,
through the Marie Curie Initial Training Network CosmoComp
(PITN-GA-2009-238356).
CSF acknowledges a Royal Society Wolfson Research Merit
Award and ERC Advanced Investigator grant 267291 COSMIWAY. This work
was supported in part by an STFC rolling grant to the Institute for
Computational Cosmology of Durham University. 
\appendix

\section{The Projected NFW Profile with background subtraction}
\label{appendix:a}
The NFW density profile \citep{nfw96,nfw97} is
\begin{equation}
   \rho(r)=\frac{\delta_{\rm c}\rho_{\rm c}}
   {(r/r_{\rm s})(1+r/r_{\rm s})^2},
\end{equation}
where $\rho_{\rm c}$ is the critical density, $\delta_{\rm c}$ is the 
characteristic overdensity of the halo and $r_{\rm s}$ is a
characteristic scale length. Conventionally, the scale length
is specified in terms of a concentration defined as $c=r_{\rm
 200}/r_{\rm s}$, where the $r_{200}$ radius is defined as the
radius at which the mean interior density is $200\rho_{\rm c}$.
With these definitions it follows that
\begin{equation}
   \delta_c=\frac{200}{3}\frac{c^3}{\ln(1+c)-c/(1+c)}.
\end{equation}
We can integrate along a line of sight to obtain the
projected surface mass density
\begin{equation}
   \Sigma(R)=2\delta_{\rm c}\rho_{\rm c} r_{\rm s}\int_R^{\infty}
   \frac{1}{\sqrt{r^2-R^2}\, (r/r_{\rm s})\, (1+r/r_{\rm s})^2}{\rm d}r,
\end{equation}
where $R$ is the projected distance from the centre of the halo.
This integral can be solved analytically 
\citep{bar96} and  expressed as
\begin{eqnarray}
   \Sigma(x)=\left\{
   \begin{array}{ll}
	\frac{\displaystyle{2\delta_{\rm c}\rho_{\rm c} r_{\rm s}
}}{\displaystyle{(x^2-1)}}
	\left [\displaystyle{1- \frac{2}{\sqrt{(1-x^2)}}{\rm arctanh}\sqrt{
	\frac{1-x}{1+x}}}\right ]& \\
	\hspace*{5cm}x<1,& \\
	\displaystyle{\frac{2\delta_{\rm c}\rho_{\rm c}r_{\rm s}}{3}}\hspace*{3.9cm} x=1,& \\
	&\\
	\displaystyle{\frac{2\delta_{\rm c}\rho_{\rm c}r_{\rm s}}{(1-x^2)}}
	\left [\displaystyle{1-\frac{2}{\sqrt{(x^2-1)}}{\rm arctan}\sqrt{
	\frac{x-1}{1+x}}}\right ]& \\
	\hspace*{5cm}x>1, &
   \end{array}
   \right.
\end{eqnarray}
where $x=R/r_s$.
In our profile measurements, we remove the 
contamination of interlopers by subtracting the 
mean density of galaxies in an outer annulus. This outer annulus
will also contain genuine satellites that are in the outer annulus
of the density profile.
Hence, to compare fairly  with the measured profiles,
we should apply the same background subtraction
process to the projected NFW profile.  We
denote the resulting background-subtracted projected NFW 
profile as
\begin{equation}
   \hat\Sigma(x)=\Sigma(x)- \frac{2 r_s^2}{3 r_{200}^2 }  
\int_{x_{200}}^{2x_{200}} \Sigma(x)\, x \,{\rm d}x.
\label{eqn:nfw}
\end{equation}

\begin{figure}
   \includegraphics[width=84mm]{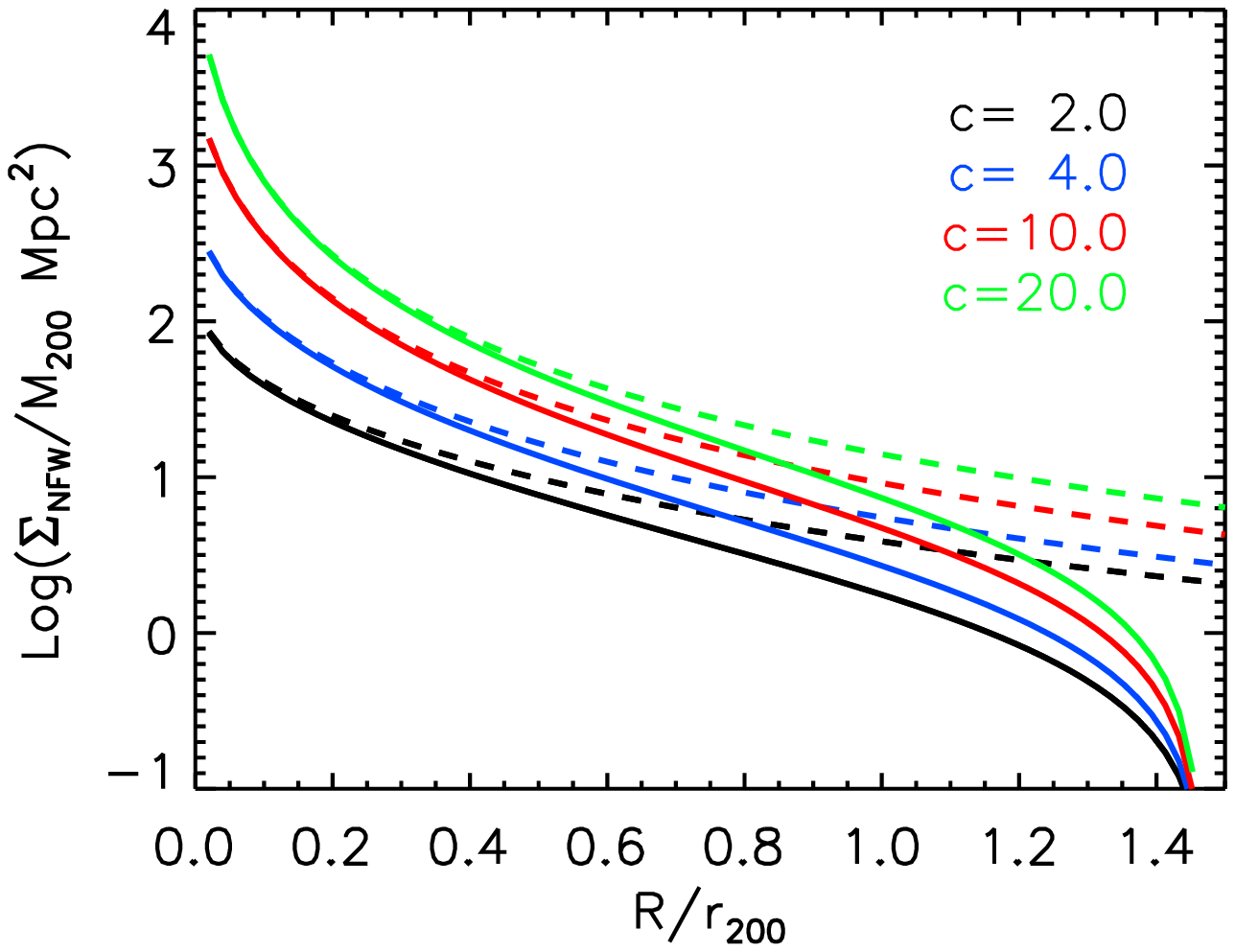}
	     \caption{A comparison of projected NFW profiles 
(dashed lines) and background-subtracted projected NFW profiles 
	     (solid lines) for different values of the concentration,
	     $c$. The projected radius, $R$, is expressed in units of 
           the $r_{200}$ radius, $r_{200}$.}
	     \label{fig:app}
\end{figure}

These background-subtracted profiles are compared to their
unsubtracted counterparts in Fig.~\ref{fig:app}. The subtracted profiles tend to
zero at projected $R/r_{200} \sim 1.4$.

\section{Comparison of results from DR8 with DR7}
\label{appendix:b}

\begin{figure}
   \includegraphics[width=84mm]{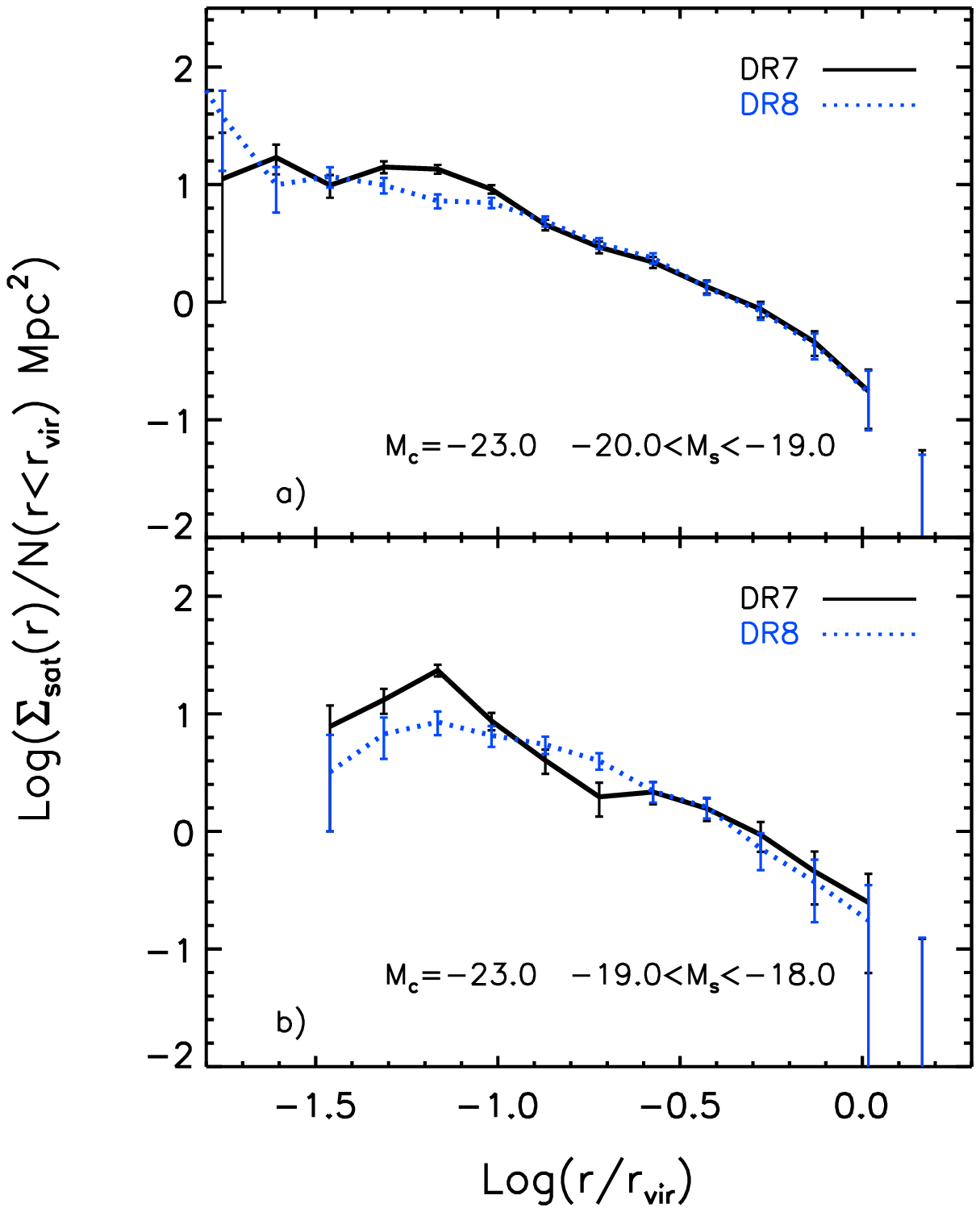}
	     \caption{A comparison of profiles based on SDSS
	     DR8 and DR7 for a) more luminous satellites, and b) less
	     luminous satellites. The black (solid) lines are the profiles
	     from DR7. The blue (dotted) lines are the profiles
	     from DR8. A cut at 0.5 times the Petrosian
	     $R_{90}$ radius is used to highlight the difference.}
	     \label{fig:diff_dr7_dr8}
\end{figure}

We have performed our measurement of satellite number density profiles
for both the SDSS DR8 and DR7. This helps to quantify the impact on
the number density profiles of the different sky-subtraction
algorithms used to define these galaxy catalogues. Looking at images
of some of our primary galaxies in DR7, there were occasions when
close-in satellite galaxies existed that were not present in DR8. The
suggestion is that these are spurious fragments of the primary
galaxy itself, following inexact subtraction of the background sky
level. One would expect such a problem to be worse for lower
luminosity satellites and also if the inner radius cut to be
considered is reduced beyond our default 1.5 times the Petrosian $R_{90}$.
Fig.~\ref{fig:diff_dr7_dr8} shows some illustrative results where
satellites down to 0.5 times the Petrosian $R_{90}$ have been included
in the profiles around $\satmc=-23.0$ primaries. There is a tendency
for DR7 to have extra low luminosity satellites near to the primary,
which is not shared by DR8. This is particularly evident in the lower
panel. Furthermore, while the DR8 profile is robust to changing the
inner radius cut, the result for the low luminosity satellite profile
around bright primaries from DR7 changes significantly.
We conclude that DR7 contains more spurious fragmentation of bright
primary galaxies, and that DR8 is preferable for our study, both in
terms of the reliability of the faint galaxies and their improved photometry.

\section{Validation of the satellite search parameters}
\label{appendix:c}
\begin{figure*}
    \begin{center}
        \includegraphics[width=158mm]{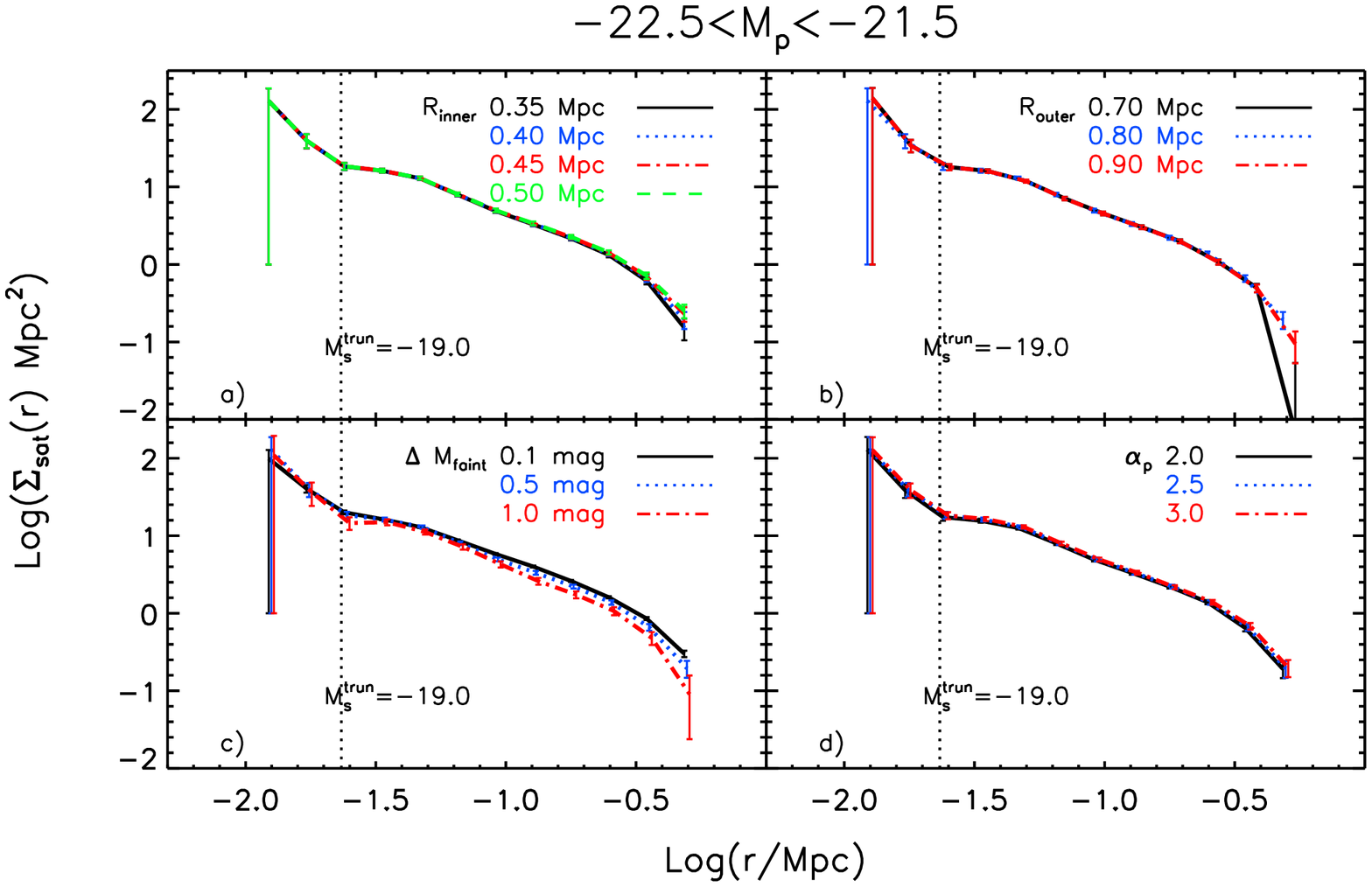}
        \caption{The effect on the estimated number density profiles of varying
        the parameters \{$\ri,\ro,\satdmf,\satap$\} from their
        default values, \{0.3 Mpc,0.6 Mpc,0.5,2.5\}, as indicated
        in the legends. Some error bars for different datasets 
    have been slightly shifted for clarity.}
        \label{fig:var}
    \end{center}
\end{figure*}

In Fig.~\ref{fig:var}, we show the effect on the estimated number density profiles 
of varying various satellite search parameters. Panel (a)
demonstrates that varying $R_{\rm inner}$ between 0.35 and 0.50 Mpc does 
not change the profiles significantly, because $R_{\rm inner}=0.35$ Mpc is
already large enough to enclose the whole satellite system for primaries
satisfying $-22.5<M_{\rm p}<-21.5$. The satellite number density
profile is similarly robust to changes in $\ro$, which is 
the outer radius for the background region, as shown in
panel (b). The next panel shows the effect of varying $\satdmf$, the
parameter used to determine if a primary is isolated. There is a very
weak variation of the profile shown in panel (c), with primaries
allowed to have neighbours with a magnitude difference as small as
$\satdmf=0.1$ having slightly more satellites than those with larger
magnitude differences to their neighbours, more in keeping with the
term isolated. 
Besides the physically motivated parameters, we also test the
parameters of the estimation method. The parameter $\satap$ helps us
to distinguish genuine satellite galaxies from background galaxies by 
excluding galaxies that are at a significantly different redshift.
Panel (d) shows that our results are insensitive to reasonable
changes in the value of $\satap$.

%\clearpage

\end{document}